\documentclass[twocolumn]{aa} 

\def\specchar#1{{\sc #1}}

\def\degree{\hbox{$^\circ$}}
\def\arcsec{\hbox{$^{\prime\prime}$}}
\def\FeI{\mbox{Fe\,\specchar{i}}}

\def\BaII{\mbox{Ba\,\specchar{ii}}}
\def\CaIIH{\mbox{Ca\,\specchar{ii}\,\,H}}       

\def\pun{\stackrel{}{\mbox{.}}}
\def\farcs{$\stackrel{\prime\prime}{\pun}$}

\usepackage{natbib}
\usepackage{graphicx}
\begin{document}
   \title{Properties of oscillatory motions in a facular region}

\author{R. Kostik\inst{1}, E. Khomenko\inst{2,3,1}}

\institute{Main Astronomical Observatory, NAS, 03680, Kyiv,
Ukraine\\ \email{kostik@mao.kiev.ua} \and Instituto de
Astrof\'{\i}sica de Canarias, 38205 La Laguna, Tenerife, Spain
\and Departamento de Astrof\'{\i}sica, Universidad de
La Laguna, 38205, La Laguna, Tenerife, Spain\\
}

\date{Received 25 of July, 2013; accepted 30 of September, 2013}

\abstract {}
{We study the properties of waves in a facular region of moderate
strength in the photosphere and  chromosphere. Our aim is to analyse
statistically the wave periods, power and phase relations as a function of
the magnetic field strength and inclination.}
{Our work is based on observations obtained at the German Vacuum Tower
Telescope (Observatorio del Teide, Tenerife) using two different instruments:
the Triple Etalon SOlar Spectrometer (TESOS), in the \BaII\ 4554 \AA\ line to
measure velocity and intensity variations through the photosphere; and,
simultaneously, the Tenerife Infrared Polarimeter (TIP-II), in the \FeI\ 1.56
$\mu$m lines to the measure the Stokes parameters and  magnetic field strength
in the lower photosphere. Additionally, we use the simultaneous broad-band
filtergrams in the \CaIIH\ line to obtain information about intensity oscillations
in the chromosphere.}
{We find several clear trends in the oscillation behaviour: (i) the period of
oscillation increases by 15--20\% with the magnetic field increasing from 500
to 1500 G; (ii) the temperature--velocity phase shifts show a strikingly
different distribution in the facular region compared to the quiet region, a
significant number of cases in the range from $-180^\circ$ to 180$^\circ$ is
detected in the facula. (iii) the most powerful chromospheric \CaIIH\
intensity oscillations are observed  at locations with strong magnetic fields
(1.3--1.5 kG) inclined by 10--12 degrees, as a result of upward propagating
waves with rather small phase speeds, and temperature--velocity phase shifts
between 0$^\circ$ and $90^\circ$; (iv) the power of the photospheric velocity
oscillations from the \BaII\ line increases linearly with decreasing magnetic
field inclination, reaching its maximum at strong field locations.}
{}


\maketitle

\section{Introduction}

There are at least two important reasons to study waves in magnetized solar
regions: (i) the magnetic field modifies the properties of waves, and this
provides information on the conditions of the atmosphere where these waves
propagate; (ii) magnetic waves may carry sufficient energy to heat the
material in the chromosphere and corona. In the recent years,
 evidence has mounted demonstrating that the magnetic field greatly influences the
properties of waves not only in active regions (sunspots and plages), but also
in quiet regions (network and internetwork) \citep[see][for a
review]{Khomenko+Irantzu2013}. Waves observed in enhanced plage, facular and
network regions share certain properties, although  detailed behavior seems to
depend on the magnetic flux contained in a particular region and is apparently
different for weaker network fields compared to stronger plage fields, as we
detail below.

Spectral and spectropolarimetric slit observations of waves in plage regions
show that the maximum of oscillation power falls at frequencies of around 3
mHz (below the temperature minimum cut-off frequency $\sim$5.2 mHz) both in
the photosphere and in the chromosphere \citep{Deubner1967, Howard1967,
Blondel1971, Woods1981, Deubner+Fleck1990, Lites1993, Centeno+etal2009,
Kobanov+Pulyaev2007, Turova2011}. The dominance of low-frequency oscillations
in the chromosphere has also been reported for waves above bright network
elements in the quiet Sun \citep{Lites1993, Krijer+etal2001,
DePontieu+etal2003, Bloomfield+etal2006, Tritschler2007, Vecchio+etal2007}.
The two-dimensional filtergram observations give richer spatial information
and demonstrate that the low-frequency 3 mHz power enhancements are located
just in the intermediate surroundings of network magnetic elements at both
layers \citep{Kontogiannis2010a}. In constrast, the high-frequency 5--7 mHz
waves are `shadowed' in the chromosphere at the same locations and their
power is reduced \citep{Krijer+etal2001, Judge+etal2001, Finsterle2004b,
Finsterle2004a, Reardon2009}.

The chromospheric power distribution in stronger-field plage and facular
areas is less clear. Apparently, at the edges of magnetic elements, where the
field is supposed to be inclined,  low-frequency 3 mHz oscillations are
observed, while high-frequency 5 mHz waves are transmitted to the
chromosphere at the centres of magnetic elements with a predominantly vertical
field \citep{DePontieu+etal2004, DePontieu2005, Hansteen2006,
Jefferies+etal2006, McIntosh+Jefferies2006, Wijn2009, Stangalini2011}.
Nevertheless, vertically propagating low-frequency 3 mHz waves are also
detected in plage magnetic elements \citep{Kobanov+Pulyaev2007,
Centeno+etal2009}. Strong complex plage areas attached to sunspot regions
show the dominance of the high-frequency power at 5--7 mHz (enhanced up to
60\% compared to the quiet Sun) both in the photosphere and in the
chromosphere \citep{Brown+etal1992, Braun+etal1992, Toner+LaBonte1993,
Hindman+Brown1998, Braun+Lindsey1999, Donea+etal2000, Jain+Haber2002,
Nagashima+etal2007}. These high-frequency power enhancements are called haloes
in the literature on local helioseismic waves. The amplitudes of
low-frequency 3 mHz photospheric waves in  plages seem to be reduced
compared to the quiet Sun by as much as 20--40\% \citep{Deubner1967,
Howard1967, Blondel1971, Woods1981, Kobanov+Pulyaev2007}.

The phase shifts between the velocity oscillations at different heights,
$\phi(V,V)$, and between the intensity and velocity oscillations at the same
height, $\phi(I,V)$, are much less investigated than the power distributions.
Older slit observations provide the $\phi(I,V)$ phases that are only slightly
different (on average) between the magnetic and non-magnetic areas.
The magnitudes and signs of the $\phi(I,V)$ phase shifts would correspond to
evanescent adiabatic acoustic-gravity waves, if one neglects the magnetic
nature of these waves \citep{Lites1993, Mein+Mein1976, Deubner1990,
Lites+etal1982a}. Larger $\phi(V,V)$ phase shifts are found for 3 mHz waves
at the network borders compared to cell centres \citep{Lites+etal1982a,
Deubner+Fleck1990}. The network `shadow' areas show $\phi(V,V)$ phase
shifts indicative of upward propagating waves at 5 mHz, but a mixture of
positive and negative phase shifts for lower-frequency waves at 3 mHz.

It is clear from all above that all the basic wave characteristics, such as
amplitudes, phases and dominant frequencies, depend on the magnetic structure
of the region where they propagate. To get a complete picture of the wave
behaviour in plage and network magnetic elements, one needs information about
the magnetic field vector at  photospheric and chromospheric heights.
While many studies of the above mentioned  lack any magnetic information, others
use photospheric magnetograms or different forms of the magnetic field
extrapolation from photospheric magnetograms to obtain the magnetic field
vector \citep{Deubner1967, DePontieu+etal2004, Jefferies+etal2006,
Chitta2012,  Vecchio+etal2007, Vecchio2009, Kontogiannis2010a,
Kontogiannis2010b}. Spectropolarimetric measurements of the magnetic field
vector in plage and network wave studies are scarce
\citep{Kobanov+Pulyaev2007, Centeno+etal2009, Stangalini2011}. The later work
of \citet{Stangalini2011} is a great improvement in this respect because of
the 2D field of view used, rather than just one slit. However, these authors
concentrate their analysis on a relatively strong pore. The behaviour of waves
in weaker structures has not, to our knowledge,  been studied with a similar dataset.
This is one of the objectives of our paper.

A long-standing question addressed to observations is how
low-frequency photospheric oscillations at 3 mHz reach chromospheric heights.
It is known that, under normal conditions, waves with frequencies below the
cut-off frequency ($\nu$ = 5.2 mHz) are evanescent and cannot propagate
vertically to the upper layers. Their power contribution to the total
spectrum is reduced, since only the evanescent tail is present there.
Nevertheless, as follows from the numerous observational detections mentioned
above, low-frequency waves with sufficient power are present in magnetized
regions.

\citet{Michalitsanos1973} showed that in an atmosphere where the magnetic
field is inclined by $\gamma$ degrees with respect to the gravity vector, the
cut-off frequency is effectively lowered to a value $\omega_c\cos\gamma$,
which makes possible the propagation of  low-frequency oscillations to the
chromosphere. This mechanism was also considered by \citet{Bel1977} and
\citet{Zhugzhda+Dzhalilov1984a}. Numerical modelling, performed by
\citet{Heggland2007, Heggland2011} confirmed its effectiveness as well.
The observations of \citet{Wijn2009} and \citet{Stangalini2011} indeed reveal the
leakage of low-frequency waves from the photosphere to the chromosphere at
the peripheral parts of rather strong magnetic elements in a plage and a pore.
\cite{Stangalini2011} find a maximum transmission of 5 mHz oscillations for
the field inclined by 15 degrees, and of 3 mHz oscillations for the field
inclined by 25 degrees.

For the efficient working of the above mechanism, a plasma $\beta$ has to be
below one already in the photosphere. Otherwise slow acoustic waves are not
forced to propagate along the inclined magnetic field and are free to
propagate in any direction. Thus, rather strong magnetic fields are required
to decrease the cut-off frequency deep in the photosphere. Nevertheless,
observations show the presence of energetic 3 mHz oscillations not only at
the periphery of plages and pores, but also in  close proximity to the
network magnetic elements with weaker and mostly vertical fields \citep[e.g.,
][]{Krijer+etal2001, Centeno+etal2009, Vecchio+etal2007, Kontogiannis2010a,
Kontogiannis2010b}.

An alternative explanation was proposed by \citet{Roberts1983} and was
followed by \citet{Centeno+etal2009}. The changes in the radiative relaxation
time of temperature fluctuations in small-scale magnetic elements may lead to
an effective change in the cut-off frequency, introducing a formally propagating
component to the wave. This mechanism allows for transmission of the 3 mHz
oscillations to the chromosphere in the vertical magnetic elements, as was
confirmed in the two-dimensional MHD simulation by
\citet{Khomenko+etal2008b}. The drawback of the latter work is in the
simplified treatment of the radiative transfer by Newton's cooling law,
compared to a very detailed treatment by \citet{Heggland2007, Heggland2011},
\citep[see the discussion in][]{Khomenko+Irantzu2013}.

From all above, it is evident that more observational and theoretical studies
are needed to clarify the relation between waves and magnetic structures in
the quiet solar regions, especially those with lower magnetic flux, such as weak
plage, network and inter-network areas. This study provides observational
analysis of such a kind. We statistically analyze the properties of waves in a
an isolated plage region of intermediate strength, observed during the
minimum of solar activity in 2007. Our observational dataset includes 2D
spectropolarimetric observations of the photospheric magnetic field vector,
together with velocity and intensity measurements through the photosphere and
the chromosphere. Below, we investigate how the different parameters of
oscillations depend on the strength and inclination of the magnetic field;
we also evaluate the contribution of waves to  energy transport from the
photosphere to the chromosphere. The same dataset was used by us to study the
convection in magnetic fields \citep{Kostik+Khomenko2012}.

\begin{figure*}
\centering
\includegraphics[width=14cm]{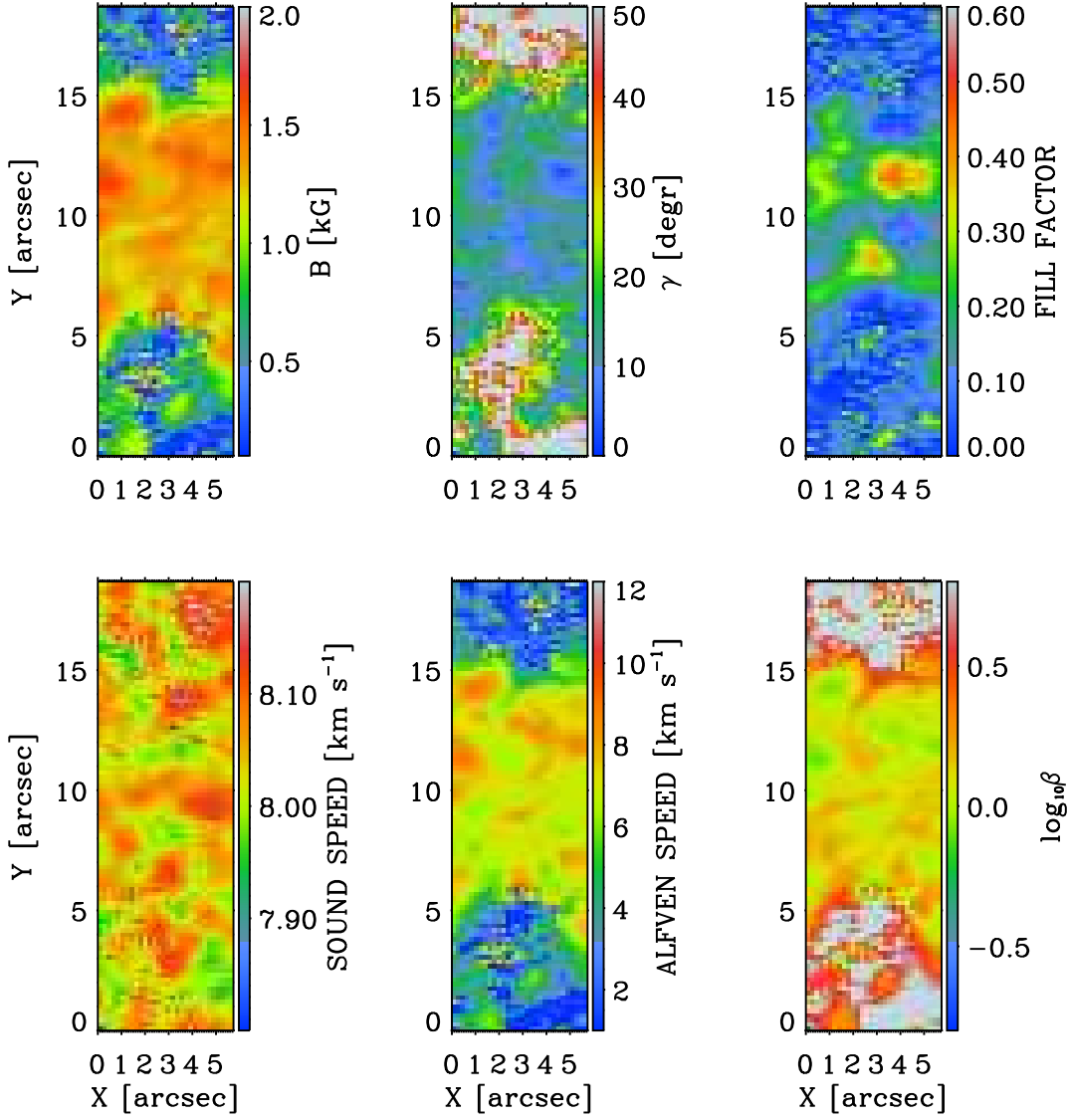}
\caption{Parameters relevant for  wave propagation, calculated from the
inversion. The upper row gives the magnetic field strength, inclination and
filling factor. The bottom row gives the acoustic speed, Alfv\'en speed and plasma
$\beta$ parameter. The maps are taken at the bottom of the photosphere at constant
log$\tau_5=-0.2$. The values of density and gas pressure are calculated
assuming hydrostatic equilibrium in each vertical column from the inversions,
and averaged over the magnetic and non-magnetic atmospheric components. }
\label{fig:inv-beta}
\end{figure*}

\begin{figure*}
\centering
\includegraphics[width=14cm]{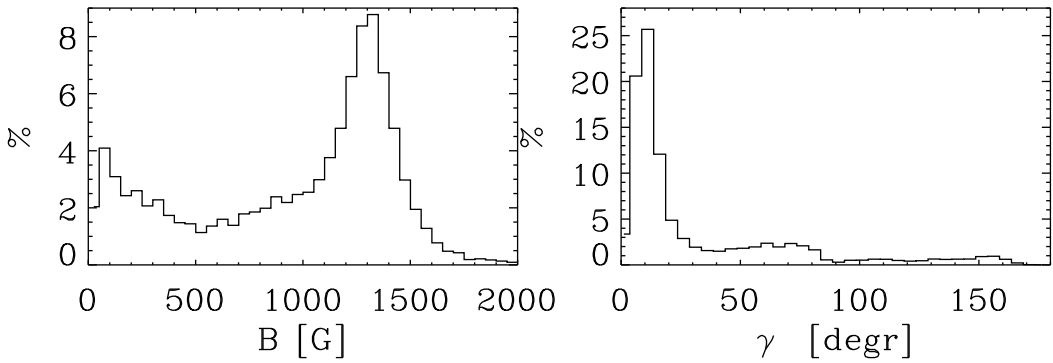}
\caption{Histogram of the magnetic field strength (left) and inclination with
respect to the line-of-sight (right) in the observed area over the five TIP
scans, as inferred from the SIR inversion.} \label{fig:b-gamma-hist}
\end{figure*}

\section{Observations and data reduction}

The observations were made on  2007 November 13  at the German Vacuum Tower
Telescope (VTT) located at the Observatorio del Teide in Iza\~na, Tenerife.
Three wavelength regions were observed simultaneously: \FeI\
$\lambda15643-15658$ \AA, \BaII\ $\lambda4554$ \AA\ and \CaIIH\ $\lambda3968$
\AA. We  used two instruments simultaneously: TIP-II \citep{Collados2007},
and TESOS \citep{Tritschler+etal2002}. On the day of the observations there
were no sunspots  present on the Sun. We have selected an isolated plage
region unattached to any sunspot group based on filtergrams in the \CaIIH\
line. The observed region was located close to the solar disk center, at
approximately S05E04. The datasets and their reduction are described in
detail in our previous paper \citep{Kostik+Khomenko2012}. We recall here only
the parameters directly relevant to the current analysis. Our dataset
consists of:
\begin{itemize}
\item Five TIP maps of Stokes spectra of \FeI\ lines at 1.56 $\mu$m,
    separated by 6 min 50 s in time and 5.\arcsec5$\times$18.\arcsec5
    in size, with a pixel size of 0\farcs185. These data were used to
    recover the magnetic field strength and inclination by means of SIR
    inversion.
\item Time series of \BaII\ 4554 \AA\ line spectra over a
    5.\arcsec5$\times$18.\arcsec5 area, co-spatial with TIP, with a
    temporal cadence of 25.6 s, pixel size of  0\farcs089 and a duration
    of 34 min 41 s. We used the \BaII\ spectra to calculate the velocity
    and intensity fluctuations at 14 levels of the line profile,
    corresponding to 14 heights in the atmosphere, by means of
    $\lambda$-meter technique. We use the convention of positive velocity
    is an upflow. Note that, despite the consecutive points of the \BaII\
    line profile were not measures simultaneously, the $\lambda$-meter
    velocities and intensities at each of the 14 levels correspond
    approximately to the same time moment, since the technique averages
    blue-wing and red-wing values.
\item Time series of \CaIIH\ filtergrams over
    5.\arcsec5$\times$18.\arcsec5 area, co-spatial with TIP and TESOS,
    with a temporal cadence of 4.93 s, pixel size of 0\farcs123 and the
    same duration as the TIP and TESOS series. This gives us intensity
    fluctuations in the low chromosphere.
\end{itemize}

The signal-to-noise ratio of Stokes profiles in the TIP data was around $2.5
\times 10^{-4}I_c$. The amplitudes of the polarization signal in the observed
area were rather strong, reaching a maximum of $\sim 0.03 I_c$ in Stokes $V$
and $0.01-0.02 I_c$ in Stokes $Q$ and $U$. The areas with the largest signal
in Stokes $V$ correspond to a local darkening in the Stokes $I$ continuum
intensity.

Stokes parameters of the \FeI\ 15648 and 15652 \AA\ lines were inverted using
the SIR inversion code \citep[Stokes Inversion based on response functions,
see][]{RuizCobo+delToroIniesta1992}. We fitted all four Stokes parameters of
the two \FeI\ lines.
We slightly changed the inversion procedure compared to
\citet{Kostik+Khomenko2012} and assumed a two-component model (one magnetic
and one non-magnetic component), with a filling factor as a free parameter.
The temperatures and line-of-sight velocities of both atmospheres were
allowed to vary independently. The magnetic field strength and orientation
were assumed constant with height. Since the Stokes signals in the observed
area are rather strong, they allow for robust fits.
The mean error in the magnetic field strength and inclination, as given by
the inversion code, is around 80 G and 7\degree\ in the strong field pixels
with more vertical field ($B>0.8$ kG, $\gamma<20$\degree). In the weaker
field pixels ($B<0.8$ kG), the error increases up to 110 G and 12\degree.
Figure~\ref{fig:inv-beta} gives some parameters resulting from the inversion
for the first TIP map for the lower photosphere at $\log \tau_5=-0.2$.
The magnetic field in the area reaches up to 2 kG. The filling factor of the
magnetic atmosphere is up to 40--60\% in the strongest patches. The
inclination map shows coherent structures, with a more vertical field at the
centres of stronger magnetic patches (in the central part of the map),
inclined by 15--20 degrees in the surrounding area. At the peripheral parts of our
observed region, the inclination becomes progressively higher, up to 40--50
degrees. There, the magnetic signal is not so strong.
Figure~\ref{fig:b-gamma-hist} summarizes these results over all five maps. It
shows that the magnetic field has a distribution with two dominant peaks, one
in the weaker fields (with a linear decrease towards the larger fields) and
another  at 1.3 kG for the magnetic elements in the plage region. The
distribution of inclinations peaks at small values (0--20 degrees), as was
already evident from Figure~\ref{fig:inv-beta}.

The bottom row of Figure~\ref{fig:inv-beta} shows maps of  parameters
calculated from the inversions that are directly relevant for the wave
propagation, i.e.\ the sound and Alfv\'en speeds, $c_s$ and $v_a$, and the plasma
$\beta$ parameter, approximated as $\beta=c_s^2/v_a^2$. To get gas pressure
and density from the inversions we assumed hydrostatic equilibrium in every
vertical column and then averaged the pressure and density values over the
magnetic and non-magnetic components with a corresponding filling factor.
The sound speed map shows granulation structure with the larger values in the
hotter granular areas, although the range of variation of this parameter is not
so strong (7.8--8.2 km s$^{-1}$). The Alfv\'en speed at this $\log\tau_5$ is
significantly larger in the central part of the map with the stronger
magnetic patch. There, the values vary over a larger range than for the
acoustic speed (between 6--10 km s$^{-1}$). The parameter $\beta \geq 1$ over
the central part of the magnetic patch. Below $\log\tau_5=-0.2$ the values of
$\beta$ are greater than unity, meaning that the atmosphere is essentially
gas-pressure dominated. Above $\log\tau_5=-0.2$ the plasma $\beta$ becomes
smaller than one, and the atmosphere is magnetically dominated. Note that the
above are only estimates, taking into account all the approximations and
uncertainties in the inversion (such as, for example, constant magnetic field with
height). Nevertheless, it gives us some idea of the distribution of forces
in the observed region, showing that the atmosphere is essentially
magnetically dominated from the lower photosphere upwards. This defines the
properties of the waves.

 As in our previous study \citep{Kostik+Khomenko2012} we have split the
variations of velocity and intensity from the \BaII\ line profiles into two,
oscillatory and convective, components based on the $k-\omega$ diagram. Below
we focus our study on the oscillatory component.

\begin{figure*}
\centering
\includegraphics[width=18cm]{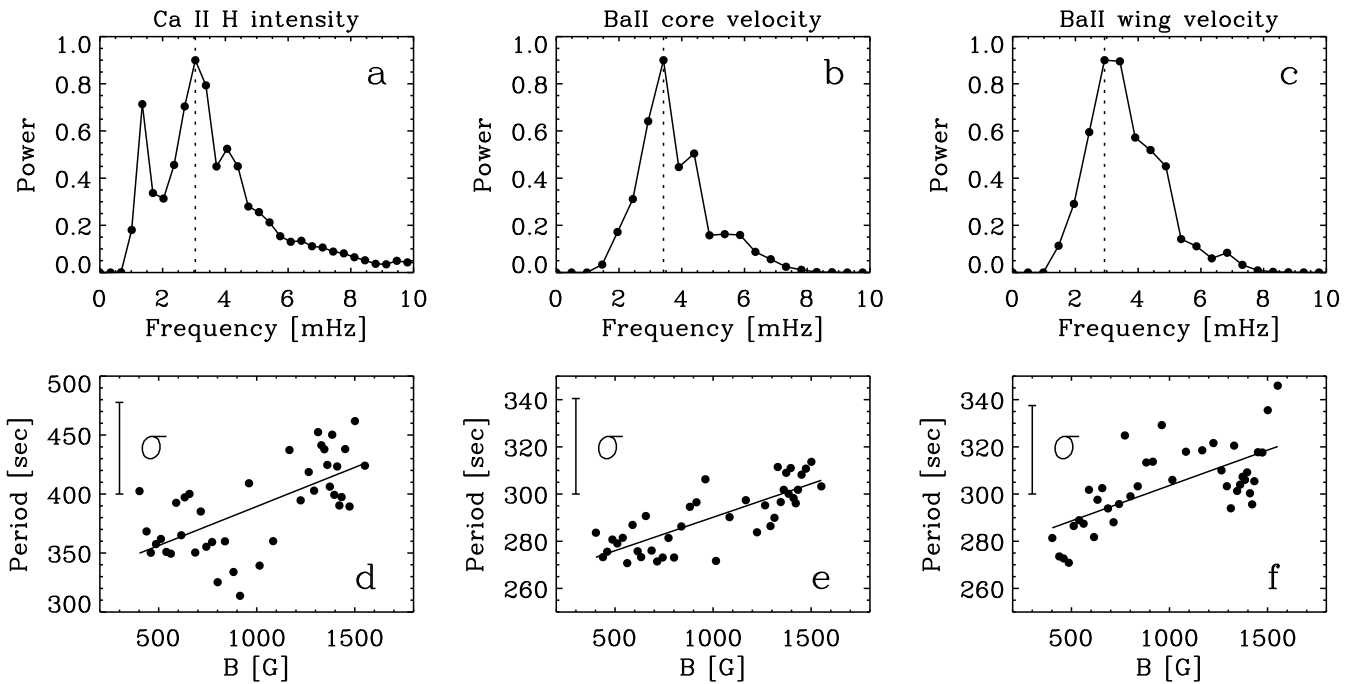}
\caption{Space-averaged power spectra (upper row, a--c) and dominant period
of oscillations as a function of the photospheric magnetic field strength
(bottom row, d--e). Left column: intensity oscillations in \CaIIH\
filtergrams (a, d); Middle column: velocity oscillations measured at the
\BaII\ line core (b, e); Right column: velocity oscillations measured at the
\BaII\ line wing (c, f). } \label{fig:period-b}
\end{figure*}

\section{Results of observations}

Figure~\ref{fig:period-b} shows the power spectrum, and the dominant periods
of oscillations at three representative heights: the low photosphere, from
the $\lambda$-meter velocities in the $\BaII$ line wings (rightmost panels);
the upper photosphere from the $\lambda$-meter velocities in the $\BaII$ line
core (middle panels); and the low chromosphere, from the \CaIIH\ intensity
variations. The power spectra are averaged over the all the spatial points
irrespective of the magnetic signal (5.5\arcsec $\times$ 18.5\arcsec). The
power spectra have a broad distribution with peaks around 3--3.5 mHz. Some
secondary peaks at lower and high frequencies are also present at all
heights. The most prominent one is at lower frequencies, 1.3 mHz in the
\CaIIH\ data. The dominant period of oscillations shows a clear dependence on
the photospheric magnetic field strength. It increases by 30--60 s with $B$
from $\langle B \rangle=500$ G to $\langle B \rangle=1500$ G at all three
representative heights. We did not take into account a possible inclined
propagation when calculating the relation between the dominant period and the
underlying photospheric magnetic field. The periods and the magnetic field
strengths are measured at the same spatial pixels. Our results mean that, on
average, upper photospheric and chromospheric oscillations in a plage appear
with larger periods just above the locations with stronger magnetic field.
Unfortunately, our dataset does not allow us to determine the magnetic field
in the upper photosphere, which means that we are unable to determine whether
the dependence between the dominant period and $B$ would remain the same for
the magnetic field measured higher up.

\begin{figure*}
\centering
\includegraphics[width=18cm]{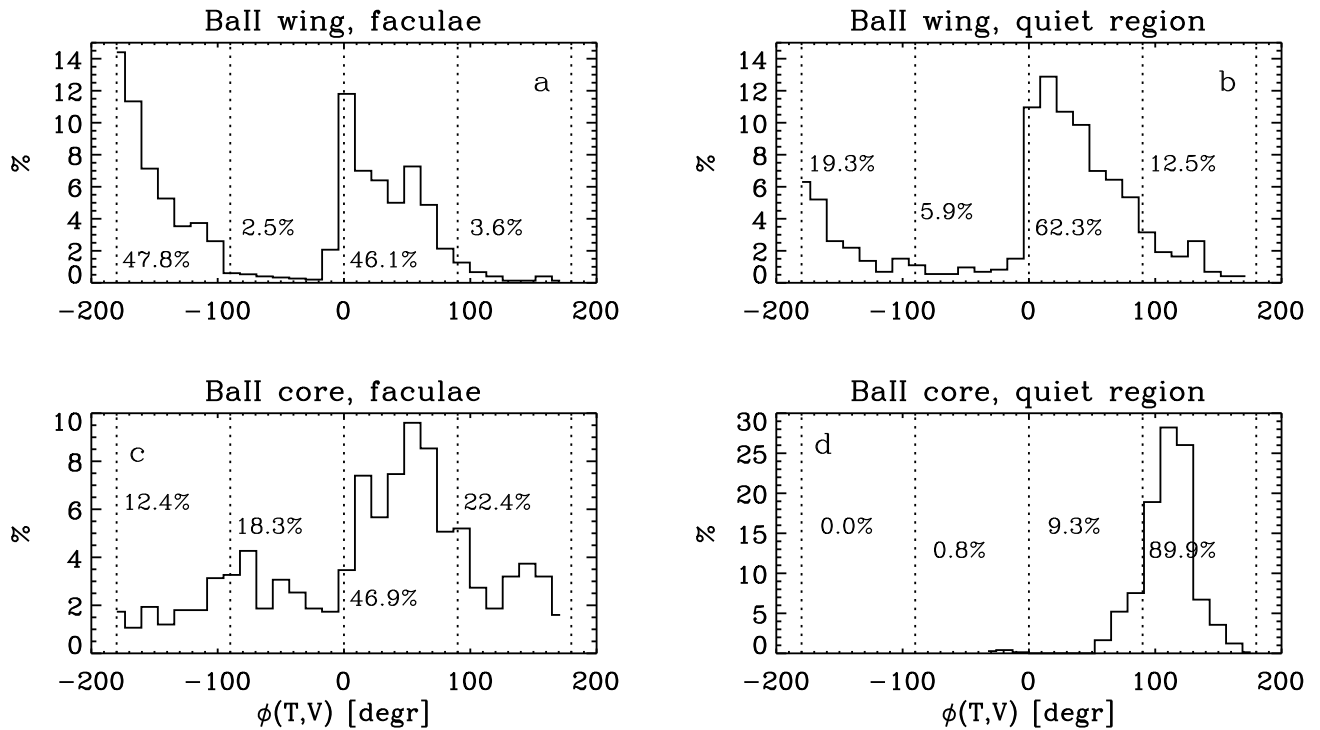}
\caption{Histograms of the phase shifts between the temperature (in
anti-phase with intensity) and velocity oscillations $\phi(T, V)$ for the
bottom photosphere (upper row), and the upper photosphere (bottom row). The
panels on the left show the results for the facular region analysed in this
paper. The panels on the right show a similar result, but for the quiet
region data from \cite{Kostik+etal2009}.} \label{fig:phi-tv}
\end{figure*}

Phase differences between oscillating parameters provide a lot of useful
information about the wave types and their propagation properties (e.g.\
adiabaticity or energy losses). We have calculated the phase shifts between
the $\lambda$-meter intensity and velocity oscillations for the \BaII\ line
at two representative levels, one near the line core, corresponding to the
upper photosphere; and one in the far wing, corresponding to the lower
photosphere. For a better interpretation of this phase shift, and for
comparison with theoretical models, we have converted the phase of the
intensity oscillations into the phase of temperature oscillations. For that,
we based on the following qualitative argument. Since we deal with a line of
a singly ionized element, an increase in temperature leads to an increase in
the number of atoms in the ionized state absorbing the radiation. The line
becomes deeper and its intensity decreases, i.e.\ the temperature and intensity
oscillations are 180 degrees out of phase. Keeping this in mind,
Figure~\ref{fig:phi-tv} presents the histograms of the distribution of the
phase shifts between temperature and velocity (positive velocity direction is
toward the observer), $\phi(T,V)$. The sign convention is that positive
$\phi(T,V)$ mean that the oscillations of temperature lead those of velocity;
the negative sign is for the opposite.

The phase shifts are frequency-dependent quantities. The $\phi(T,V)$ phase
shifts in Fig.~\ref{fig:phi-tv} are calculated at the dominant frequencies of
the oscillations and are different in each spatial pixel. We used the whole
field of view. The range of variation of frequencies is $\nu=2.90 - 3.91$ mHz
(periods between 256--341 s, see Fig.~\ref{fig:period-b}) depending on the
location.
We split the whole range of $\phi(T,V)$ values in Fig.~\ref{fig:phi-tv} into
four domains: ($-180^\circ\div-90^\circ$); ($-90^\circ\div 0^\circ$);
($0^\circ\div+90^\circ$); and ($+90^\circ\div+180^\circ$).
If there were no magnetic field in the observed region, these four domains
would correspond to a different propagation behaviour of the waves, according to
the well-developed theory of non-adiabatic acoustic-gravity waves in a
stratified atmosphere \citep{Whitney1958, Noyes+Leighton1963, Holweger1975}.
The values of  $\phi(T, V)$ between $0^\circ\div+90^\circ$ would mean upward
propagating waves, $\phi(T, V)$ between $-180^\circ\div-90^\circ$  would mean
downward propagating waves, and $\phi(T, V)$ between
$+90^\circ\div+180^\circ$ would be evanescent non-propagating waves. This
simple division is not justified once  magnetic field is introduced into
the atmosphere. Later in the paper, we investigate the correlation between the
magnitude and sign of $\phi(T, V)$ shift and the direction of the wave
propagation. We keep the division into the sub-domains since the $\phi(T, V)$
histograms naturally split into these four regimes.

The two upper panels of Fig.~\ref{fig:phi-tv} show  histograms of the
$\phi(T, V)$ phase shift for the oscillations in the \BaII\ line wing (low
photosphere). The two bottom panels are for the oscillations at the \BaII\
line core (upper photosphere). We compare our observations in the facular
region with another dataset of a similar kind, taken in the \BaII\ line at
the quiet solar disc centre with no detectable magnetic activity (panels on
the right). The latter dataset was taken in June 2004 at the VTT and was
used in our earlier work \citep{Kostik+etal2009}. The details of the
observations can be found in that paper.

The histograms of the $\phi(T, V)$ shift in the quiet region (right panels of
Fig.~\ref{fig:phi-tv}) allow for an easier interpretation. At the bottom
photosphere, the maximum number of cases falls into two domains:
($-180^\circ\div-90^\circ$) with 19.3\% of cases, and
($0^\circ\div+90^\circ$) with 62.3\% of cases. According to the theory of
acoustic-gravity waves, this can be interpreted as if most of the waves are
propagating upwards in the deep photosphere. Higher up (bottom right panel),
$\phi(T, V)$ shifts; 89.9\% of all points fall within the domain
($+90^\circ\div+180^\circ$), corresponding to evanescent waves. This result
is also easy to understand, since we consider waves in the 3 mHz frequency
range, which are evanescent under normal conditions in the upper photosphere.
The values of $\phi(T, V)$ shift between ($+90^\circ\div+180^\circ$) fit well
into the range of measurements of other authors \citep[see, e.g.\ the review
of ][]{Deubner1990}.

The magnetic facular region shows a clearly different picture. In the lower
photosphere (upper left panel of Fig.~\ref{fig:phi-tv}), most of the measured
$\phi(T, V)$ values fit into the same two domains as in the quiet region
($-180^\circ\div-90^\circ$ and $0^\circ\div+90^\circ$), but the relative
weight of the domains is different. While in the quiet region, most of the
area provides $\phi(T, V)$ between $0^\circ\div+90^\circ$, in the magnetic
region the distribution between both domains is almost half by half. In
addition, there are fewer points in the `evanescent' domain
($+90^\circ\div+180^\circ$) in the facular region (3.6\% vs.\ 12.5\%).

Even more striking differences between the quiet and the magnetic region are
observed in the upper photosphere. The histogram of the facular region
shows a significant number of points within all four phase domains, unlike in
the quiet region. The maximum number of cases (46.9\%) falls within the
domain ($0^\circ\div+90^\circ$), while only 9.3\% of points in the quiet
region belong there. The number of points in this domain remains almost the
same both in the lower and upper photosphere in the facular region
($\approx$46\%). The number of points in the domains
($+90^\circ\div+180^\circ$) and  ($-90^\circ\div0^\circ$) has significantly
increased from the bottom to the top of the photosphere. Note that the $\phi(T, V)$
phase shifts between  $-90^\circ\div0^\circ$ are not present in the upper
layers of the quiet photosphere. Such phase differences are not possible for
the acoustic gravity waves according to the theoretical calculations.
Since the difference between both regions lies essentially in the presence of
the magnetic field, we suppose that waves showing $\phi(T, V)$ between
$-90^\circ$ and 0$^\circ$ in the upper photosphere are some type of
magneto-acoustic waves. Note that these types of waves are almost absent
at the bottom of the photosphere of the facular region where the plasma $\beta$ is
around or above unity (gas-pressure dominated region), see
Fig~\ref{fig:inv-beta}. Fig.~\ref{fig:phi-tv} clearly shows that the nature
of the waves propagating in the quiet and magnetized areas is fundamentally
different. The interpretation of these results in terms of the wave modes is
not straightforward and must rely on theoretical modelling, taking into
account a particular magnetic field distribution in the observed region.

\begin{figure*}
\centering
\includegraphics[width=12cm]{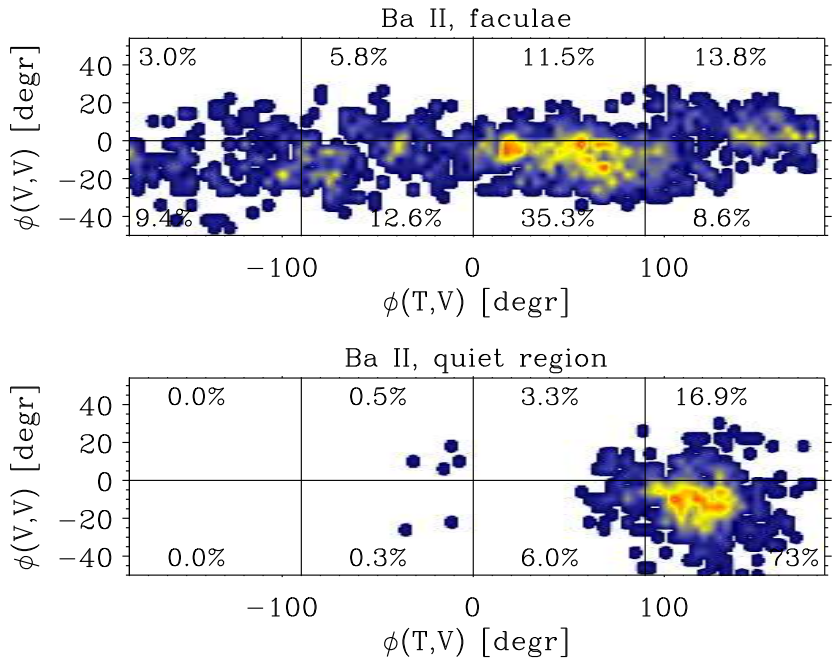}
\caption{Phase shift between the velocities at the bottom and top of the photosphere
$\phi(V,V)$ as a function of the phase shift between the velocity and
intensity at the upper photosphere $\phi(V,T)$ for the plage area (upper row)
and for the quiet area (bottom row). The yellow and red colours denote a greater
density of points in the distribution. Negative velocity--velocity shifts,
$\phi(V,V)$, mean upward waves.} \label{fig:phi-tv-vv}
\end{figure*}

Since in general, one cannot expect  the sign and  magnitude of the
$\phi(V,T)$ phase shift to give the direction of  wave propagation, once the
magnetic field is included into consideration, we found it interesting to
investigate the statistics between the measured $\phi(V,T)$ shifts and the
phase shifts between the velocity oscillations at two different heights from
the \BaII\ line core and wings, $\phi(V,V)$. These statistics are given in
Figure~\ref{fig:phi-tv-vv}. We compare the results for the facular region
(upper panel) and the quiet region (bottom panel).
In the quiet region, most of the points are located in the evanescent domain,
$\phi(V,T)$ between ($90^\circ\div180^\circ$). Among all the points, 73\% show
negative velocity phase shifts with most frequent values above $-20^\circ$.
Thus, the phase of the wave changes only slightly with height, compatible
with their evanescent behaviour. The values of the $\phi(V, V)$ shifts
measured in the quiet region are compatible with those from the earlier
studies \citep[e.g.,][]{Deubner+Fleck1989, Fleck+Deubner1989,
Deubner+Fleck+Marmolino+Severino1990, Deubner+Fleck+Schmitz+Straus1992,
Khomenko+Kostik+Shchukina2001}.

In the facular region, about a half of the points are located in the domain
$\phi(V,T)$ between ($0^\circ\div90^\circ$). The velocity phase shifts
corresponding to this domain are mostly negative, but are smaller in absolute
value compared to the quiet region. This can be interpreted as lower upward
propagation speeds of waves in the magnetic region. There is also a
significant number of points in the other three domains. The areas with
$\phi(V,T)$ belonging to ($-180^\circ\div-90^\circ$) give negative velocity
phase shifts in most cases with larger absolute values up to 40$^\circ$ and a
large scatter. These waves are upward propagating with larger phase speeds.
On the contrary, waves with $\phi(V,T)$ belonging to ($90^\circ\div180^\circ$)
show mostly positive velocity phase shifts (downward propagation) with small
phase velocities. Waves with $\phi(V,T)$ belonging to
($-90^\circ\div0^\circ$) show a large scatter and a slight preference for the
negative velocity phase shifts. Altogether, these results are not
straightforward to interpret and will need further studies.

\begin{figure*}
\centering
\includegraphics[width=18cm]{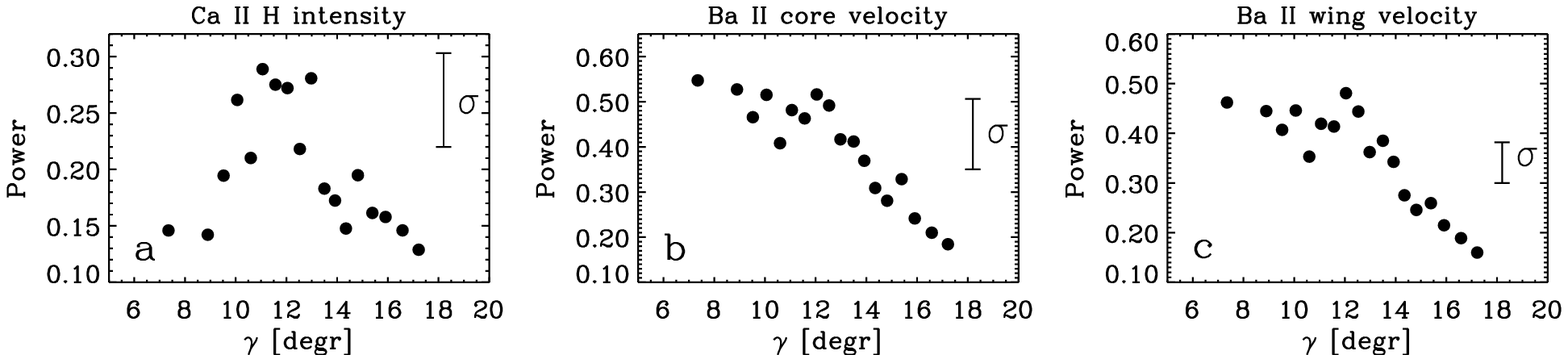}
\includegraphics[width=18cm]{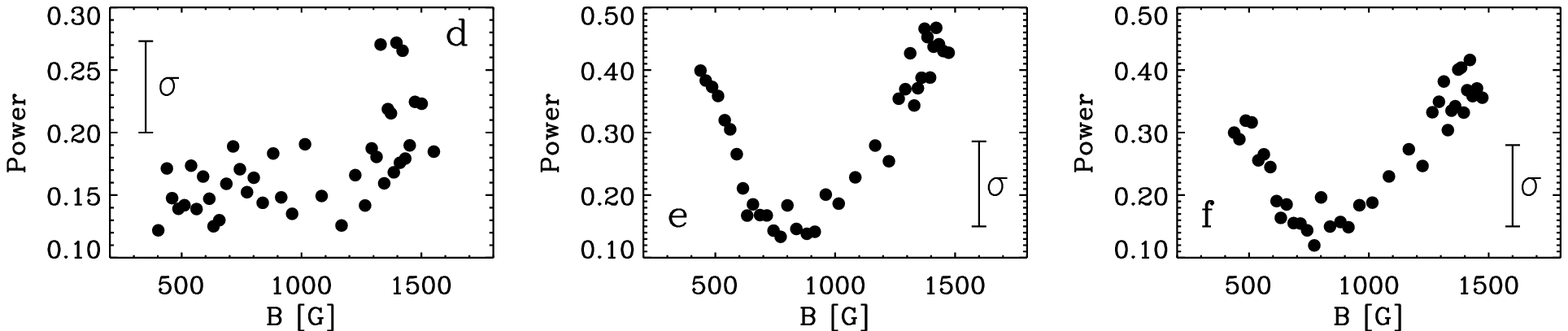}
\caption{Upper panels (a--c): power of the intensity oscillations measured
from \CaIIH\ (a); velocity oscillations at the \BaII\ core (b); and velocity
oscillations in the \BaII\ wing (c), as a function of the magnetic field
inclination in the photosphere. The values of inclination are averaged in
time over the five TIP maps. Lower panels (d--f): same, but as a function of
the photospheric magnetic field strength.} \label{fig:power-g}
\end{figure*}

\begin{figure*}
\centering
\includegraphics[width=18cm]{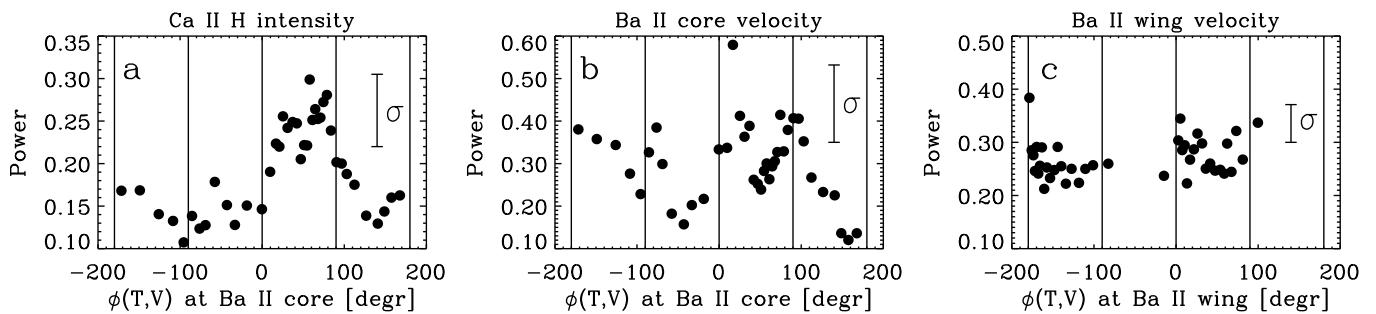}
\caption{Power of the intensity oscillations measured from \CaIIH\ (a),
\BaII\ core (b) and \BaII\ wing (c) as a function of the phase shift between
the velocity and temperature oscillations, $\phi(T, V)$. The values of
$\phi(T, V)$  are measured at the \BaII\ line core at the panels (a, b); and
in the \BaII\ line wing in  panel (c). } \label{fig:power-phi-tv}
\end{figure*}

Next, we considered the relation between the oscillation power at different
heights, and the magnetic field strength and inclination, measured in the
photosphere. Figure~\ref{fig:power-g} gives the result. The power of
oscillations is taken at the same frequency at all spatial points. As a
reference frequency we took the value of the spatially averaged dominant
frequency from Fig.~\ref{fig:period-b}, 3.41 mHz for the \BaII\ velocity
oscillations, and 3.04 mHz for the \CaIIH\ intensity oscillations. We
correlate the magnetic field strength and inclinations with the wave power at
the same spatial pixel without considering a possible inclined propagation
and spatial displacement of the wave front.

It appears that the power of the oscillations does indeed depend on the magnetic
field inclination beyond the error bar limits at all considered heights
(upper panels of Figure~\ref{fig:power-g}). The photospheric velocity from
the \BaII\ line shows an almost linear dependence: the power increases for
less inclined fields. There is a hint of another maximum of the
distribution for the inclinations between 10 and 12 degrees. Altogether, the
power of photospheric oscillations seems to be a maximum for the fields
inclined below 12 degrees.
The power distribution is different in the chromosphere. The power of \CaIIH\
intensity oscillations shows a clear maximum for the fields inclined by
10--12 degrees. This maximum can possibly be due to a reduction of the
cut-off frequency of 3 mHz oscillations because of the inclined propagation
in the low-$\beta$ regime. According to Fig.~\ref{fig:inv-beta}, the plasma
$\beta$ is below unity in most of the observed facular region from the middle
photosphere upwards. The maximum power of the photospheric oscillations at
less inclined fields must be due to some other mechanism, however.

The power of oscillations also depends on the photospheric magnetic field
strength (see the bottom panels of Figure~\ref{fig:power-g}). In the
photosphere (panels e--f), the power of the \BaII\ velocity oscillations
shows a non-linear dependence on $B$. There is a tendency for the power to
increase in the strongest field areas (1.3--1.5 kG). The power of
oscillations shows a minimum for the intermediate field strength and another
maximum for the weakest fields. The presence of these two maxima seems to be
in accordance with a hint of a two-family distribution of the power with the
inclination (panels b--c). Photospheric velocity oscillations with higher
power are observed above the regions where the field is strong  and  slightly
inclined (by 10--12 degrees). Nevertheless, there are photospheric
oscillations with a strong power above the regions with weak and less
inclined fields.
The power distribution of the chromospheric \CaIIH\ intensity oscillations
 shows only one maximum for the largest field strength. This demonstrates that
waves above the weaker field regions do not reach the chromosphere with
sufficient power.

Finally, figure~\ref{fig:power-phi-tv} shows the power of oscillations at
different heights as a function of temperature--velocity phase shift $\phi(T,
V)$. There are systematic differences in the power of waves at the four
sub-domains of the $\phi(T, V)$ values. In the lower photosphere (panel $c$),
the points are almost equally distributed between two domains with $\phi(T,
V)$ in the range ($-180^\circ\div-90^\circ$) and ($0^\circ\div90^\circ$). The
power is the same in both domains within the error bar limits. The
distribution of points is a replica of the $\phi(T, V)$ histogram in
Fig.~\ref{fig:phi-tv}, with 94\% of all cases falling into these two
categories.

The power histogram for the upper photosphere (panel $b$) seems to be an
intermediate situation between tthat for the lower photosphere ($c$) and for the
chromosphere ($a$). The power of  the chromospheric \CaIIH\ intensity
oscillations  shows a clear maximum for the phase shifts in the domain
($0^\circ\div90^\circ$). Not only is there a large number of cases of
oscillations with this phase shift detected in the upper layers but
their power is also enhanced. The interpretation of this result in terms of the
wave propagation needs theoretical modelling.

Summarizing all the above, we find that in the observed facular region, the most
powerful waves reach the chromosphere at locations with the strong magnetic
field inclined by 10--12 degrees as a result of  upward propagation with
rather small phase speeds. These waves have a preference for the
temperature--velocity phase shifts in the range $0^\circ\div90^\circ$. The
periods of the chromospheric facular oscillations are in the 5 min range,
increasing with magnetic field strength.

\begin{figure}
\centering
\includegraphics[width=9cm]{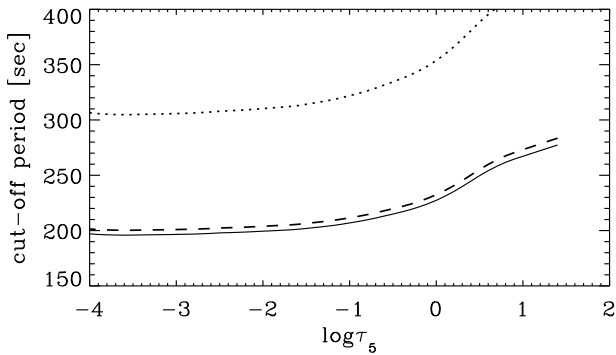}
\caption{Solid line: average cut-off period in the observed facular region as a function of
 optical depth $\log\tau_5$. The dashed and dotted line give the cut-off period for the magnetic
field inclined by 12$^\circ$ and 50$^\circ$, correspondingly.} \label{fig:cut-off}
\end{figure}

\section{Discussion}

The results of our study suggest that waves have a dominant period of around 5
minutes in the observed facular region of intermediate strength
(Fig.~\ref{fig:period-b}). This dominant period is maintained both through
the photosphere and in the low chromosphere. This conclusion is consistent
with the previous studies of waves in enhanced network and facular regions
\citep{Deubner1967, Howard1967, Blondel1971,  Woods1981, Lites1993,
Deubner1990, Krijer+etal2001, Centeno+etal2006, Kobanov+Pulyaev2007,
Centeno+etal2009}. At all the observed heights the dominant period of
oscillations increases by 15--20\% with the magnetic field increasing from 500 to
1500 G. Our results give yet further evidence that the long-period waves can
propagate with sufficient power to the upper layers in magnetic regions.

One of the possible explanations of this effect, as already mentioned in the
introduction,  would be a decrease in the acoustic cut-off frequency due to
field-aligned propagation of the slow acoustic waves in a strong magnetic field
\citep{Michalitsanos1973, Bel1977, Zhugzhda+Dzhalilov1984a, Heggland2007,
Wijn2009, Heggland2011, Stangalini2011}. This mechanism would be at work if the
plasma $\beta$ were below unity over the observed heights. The results of
spectropolarimetric inversion of our infrared Stokes data suggest that $\beta
< 1$ in the magnetic patch from the low photosphere upwards. Thus, the
reduction of the cut-off due to the field inclination should play a role in
the observed chromospheric power distribution.
We checked the effectiveness of this mechanism by plotting the power of
the \CaIIH\ oscillations as a function of photospheric magnetic field strength
and inclination for waves with periods around 5 min
(Fig.~\ref{fig:power-g}$a$ and $d$). Indeed there is a pronounced maximum of
power for the magnetic field inclined by 10--12 degrees and $B$ of 1.3--1.5
kG. In contrast, velocity oscillations in the upper photosphere from the
\BaII\ line core show a linear increase of power with decreasing magnetic
field inclination, with a hint of a secondary maximum around 10--12 degrees
(Fig.~\ref{fig:power-g}$b$). Nevertheless, the magnetic field was determined
at the bottom of photosphere, so its inclination and strength may well be
different at heights of formation of \CaIIH\ intensity observed by the
broad-band filter. \citet{Stangalini2011} find that a maximum of power of the
5-min oscillations falls for the magnetic field inclined by about 25 degrees,
for the magnetic region containing an isolated pore. The difference may be
due to a stronger field in the region observed by \citet{Stangalini2011} due
to a different configuration of the magnetic field and thermal structure of
the region, as well as due to the difference in heights sampled by the
observed spectral lines.

Taking the data from the inversion, we have evaluated the average cut-off
period in the observed facular region, and present it in
Figure~\ref{fig:cut-off}. The period decreases from 300 to 200 s through
the photosphere. Assuming that the field keeps its inclination with height, the
value of $\gamma=12^\circ$, would give us a dashed curve for the cut-off
period. This increase is obviously insufficient to allow for the efficient
propagation of the 5-minute waves. Only allowing for the field inclination
above 50 degrees do we reach a considerable increase in the cut-off period
(the dotted line in Fig.~\ref{fig:cut-off}). This leads to the conclusion that
either the magnetic field becomes steeply more inclined by at least
50$^\circ$ in the photosphere or there must be some other mechanism
facilitating the energy propagation of the 5-minute waves.

An alternative mechanism for the propagation of the 5-minute oscillations to
the chromosphere is the action of radiative losses, adding a formally
propagating component to the oscillations and preventing them from being
evanescent \citep{Roberts1983, Khomenko+etal2008b}.
The evanescence of the acoustic-gravity waves in the quiet Sun can be
evaluated by means of their phase relations, $\phi(T, V)$. We find that in
the upper photosphere in the quiet area most of measured $\phi(T, V)$ values
belong to the evanescent domain ($90^\circ\div180^\circ$)
(Fig.~\ref{fig:phi-tv}$d$). Analytical treatment of non-adiabatic magnetic
waves in the stratified atmosphere is significantly more complex even in the
case of constant magnetic field \citep{Babaev+Dzhalilov+Zhugzhda1995a,
Babaev+Dzhalilov+Zhugzhda1995b, Khomenko+etal2003}. It is difficult to make
theoretical predictions for the phase relations in the magnetic case. Our
data reveal a striking difference between the $\phi(T, V)$ distributions in
the quiet and facular regions (Fig.~\ref{fig:phi-tv}$c-d$). About a half of
the observed facular area shows $\phi(T, V)$ phase shifts in the domain
($0^\circ\div90^\circ$) that would correspond to the upward propagating
acoustic waves if there were no magnetic field. There is a significant number
of points with $\phi(T, V)$  in the whole interval from $-180^\circ$ to
180$^\circ$, unlike in the quiet region. These data confirm that wave
facular areas are significantly affected by the magnetic field. Nevertheless,
the theoretical interpretation of these data is not immediately apparent, and we cannot
yet arrive at any conclusions about the effectiveness of radiative losses of waves in the
magnetic region without extensive theoretical modelling.

We have checked whether there is any relation between the sign and magnitude of
the $\phi(T, V)$ phase shift and the direction of the wave propagation in the
magnetic area similar to the quiet area. We did this by correlating the
values of the $\phi(T, V)$ phase shifts with the velocity--velocity phase shifts
at different heights, $\phi(V, V)$. On average, most of the points with
$\phi(T, V)$  in the domain ($0^\circ\div90^\circ$) show an upward propagation
in velocity (Fig.~\ref{fig:phi-tv-vv}). The velocity phase shifts are rather
small, which can be interpreted as rather low phase propagation speeds of
these waves. In constrast, the $\phi(T, V)$ domain ($90^\circ\div180^\circ$)
(corresponding to evanescent acoustic-gravity waves when $B=0$), shows the
dominance of downward-propagating waves according to the sign of the $\phi(V,
V)$ shifts.

Altogether, this leads us to conclude that most of the energetic waves with periods
of around 5 minute reach the chromosphere as a result of the upward propagation
from the photosphere in the areas with slightly inclined magnetic field. The
low $\phi(V, V)$ shifts (phase speeds) are difficult to fit into this
picture, however. The propagation speed of magnetic waves is expected to be
generally larger than of the acoustic waves in the atmosphere with $v_a>c_s$,
as follows from Fig.~\ref{fig:inv-beta}. Further observational and
theoretical studies would be needed to clarify this and other remaining
issues concerning the wave propagation in solar facular regions.

\section{Conclusions}

We have examined how the presence of a magnetic field of  intermediate
strength in a facular area affects the properties of waves. We have statistically
investigated the dependence between the periods, power and phase relations of
waves, and the magnetic field strength and inclination. Our main results can
be summarized as follows:

We confirm that 5-minute oscillations with sufficient power are able to
propagate through the whole photosphere to the lower chromosphere. At all the
observed heights, the dominant period of oscillations increases by 15--20\%
with the magnetic field increasing from 500 to 1500 G.

At the bottom of the  photosphere, the phase shift between the temperature and
velocity oscillations $\phi(T,V)$ in the majority of cases falls into two
domains: ($-180^\circ\div-90^\circ$) and ($0^\circ\div90^\circ$), both in the
facular and in the quiet region. The relative weight of the points in each
domain is different in the magnetic and the quiet area. In the upper
photosphere, the distributions of  $\phi(T,V)$ in the quiet and facular area
are more significantly  different. In the quiet area, about 90\% of cases fall
into the domain ($90^\circ\div180^\circ$), corresponding to evanescent
acoustic-gravity waves. In the magnetic facular area, the histogram shows a
significant number of $\phi(T,V)$ values over the whole range, from
$-180^\circ$ to $180^\circ$.

By correlating the temperature--velocity, $\phi(T,V)$, and the
velocity--velocity, $\phi(V,V)$ phase shifts we find that waves with
$\phi(T,V)$ between ($0^\circ\div90^\circ$) show, on average, upward
propagation with rather small phase speeds. Such phase shifts are detected in
about half of all cases in the facular area. The most powerful oscillations in
the chromospheric \CaIIH\ intensity have $\phi(T,V)$ between
($0^\circ\div90^\circ$).

The maximum power of the photospheric oscillations appears in the areas with
the magnetic field inclined by less than 12 degrees. In the chromosphere,
oscillations of maximum power are observed above the areas where the
photospheric magnetic field is inclined by 10--12 degrees, and has a strength
of 1.3--1.5 kG.

\begin{acknowledgements}
This work is partially supported by the Spanish Ministry of Science through
projects AYA2010-18029 and AYA2011-24808.
\end{acknowledgements}


\end{document}